\documentclass[]{article}
\usepackage{graphicx}
\DeclareGraphicsExtensions{.pdf,.png,.jpg}
\usepackage{setspace}
\usepackage{amsmath}
\usepackage{hanging}
\usepackage[english]{babel}
\usepackage[comma]{natbib} 
\setcitestyle{notesep={:}} 
\usepackage{url}
\usepackage{ulem} 
\usepackage{hyperref}
\usepackage{amssymb,tabu}
\usepackage{wrapfig}
\usepackage{lscape}
\usepackage{rotating}
\usepackage{epstopdf}
\usepackage{booktabs}
\usepackage[margin=1.25in]{geometry}
\usepackage{authblk} 
\DeclareMathSizes{12}{30}{16}{12}
\title{{The Bootstrapped Robustness Assessment for Qualitative Comparative Analysis}}
\author{C. Ben Gibson and Burrel Vann Jr.\thanks{Corresponding author: C. Ben Gibson can be reached at cbgibson@uci.edu}}
\affil{University of California, Irvine\\
\vspace{12pt}
\vspace{12pt}
\vspace{12pt}
\vspace{12pt}
\vspace{12pt}
\vspace{12pt}
}

\begin{document}
\maketitle
\doublespace
\newpage
{{{\normalsize{\begin{center}\begin{singlespace}{\textbf{Author Biographies}}\end{singlespace}\end{center}}}}
\vspace{12pt}
\vspace{12pt}
{{{\normalsize{\begin{singlespace}
{\textbf{C. Ben Gibson}} is a graduate student of Sociology at the University of California, Irvine. He models communication processes in social media during natural disasters; develops methods for social network analysis, demography, and QCA; and has substantive interests in social epidemiology and medical sociology, with a special focus on sexual contact networks and the mental health effects of inequality within intimate networks. 
\end{singlespace}}}}
\vspace{12pt}
{{{\normalsize{\begin{singlespace}
{\textbf{Burrel Vann Jr.}} is a graduate student in sociology at the University of California, Irvine and a Ford Foundation Predoctoral Fellow. His research examines politics, social movements, inequality, drugs, and crime. His current projects investigate social structural influences on politics, with particular attention to the effect of crime and segregation on marijuana legalization, movement impacts on politician decision-making, and the consequences of movement dynamics on the pace of statewide budgetary legislation. 
\end{singlespace}}}}

\newpage

{{
{\LARGE{\begin{center}\begin{singlespace}The Bootstrapped Robustness Assessment for Qualitative Comparative Analysis\end{singlespace}\end{center}}}}

\begin{abstract}
Qualitative Comparative Analysis (QCA) has been increasingly used in recent years due to its purported construction of a middle path between case-oriented and variable-oriented methods. Despite its popularity, a key element of the method has been criticized for possibly not distinguishing random from real patterns in data, rendering its usefulness questionable. Critics of the method suggest a straightforward technique to test whether QCA will return a configuration when given random data. We adapt this technique to determine the probability that a given QCA application would return a random result. This assessment can be used as a hypothesis test for QCA, with an interpretation similar to a p-value. Using repeated applications of QCA to randomly-generated data, we first show that generally, the tendency for QCA to return spurious results is attenuated by using reasonable consistency score and configurational N thresholds; however, this varies considerably according to the basic structure of the data. Second, we suggest an application-specific assessment of QCA results, illustrated using the case of Tea Party rallies in Florida. This method, which we coin the Bootstrapped Robustness Assessment for QCA (baQCA), can provide researchers with recommendations for consistency score and configurational N thresholds.
\end{abstract}
\newpage

\section{Introduction}


Qualitative Comparative Analysis (QCA), introduced in Charles Ragin's 1987 book {\it{The Comparative Method}}, is a set of techniques designed to be a ``middle-path'' between quantitative and qualitative analysis. Ragin and others have since published many works on practical applications and extensions of the method, including the use of `fuzzy-sets,' the development of software, and many substantive applications (\citealt{dusa_and_thiem_2014,caramani_2008,ragin_2008,rihoux_and_ragin_2009,schneider_and_wagemann_2007,thiem_and_dusa_2013}). Though originally conceived has a method to analyze middle-sized samples, some herald its interpretive qualities above regression analysis in certain contexts (\citealt{katz_et_al_2005,seawright_2005,vis_2012}).


At base, QCA allows for a set-theoretic approach to social science grounded in Boolean algebra. Within this framework, QCA can identify necessary conditions and multiple explanatory combinations of conditions (``recipes'') for the value of an outcome. Using the logic of sets, QCA can provide a useful alternative for analyzing complex causation, broadening the reach of current research strategies to integrate a combinatorial, logic-of-sets framework versus the mean-based approaches in regression analysis. QCA can also be applied to middle-N sets, where regression may fail to provide robust results. This multi-method technique that enables the researcher to be ``in dialogue'' with the results through by way of `truth table' analysis. Researchers are encouraged to bring in their own case-oriented knowledge to establish causal conditions, examine the results of the truth table, investigate the placement of cases, and redefine conditions as needed. 


Despite recent popularity, QCA is not without its detractors. QCA is most often criticized for the use of the truth table and subsequent algorithm to determine a minimum configuration that derives a given outcome. The truth table summarizes agreement on an outcome value for cases with the same configuration of causal conditions; the algorithm combines many configurations in a reduced configuration that results in an outcome. Specifically, many are skeptical of its ability to successfully identify randomly-drawn data as being patternless, an important benchmark of any method. Stanley \citet{lieberson_2004} suggested that an application of QCA using random data would often lead to spurious configurations returned. Similarly, Simon \citet{hug_2013} questioned its usefulness on the grounds that it does not allow for error in hypothesis testing. Skaaning (2013) assessed the sensitivity of QCA when varying current robustness specifications. Though \citet{braumoeller_2015} embarked on a permutation test to determine which configurations returned via QCA are due to randomness, a principled test of QCA against {\it{totally}} random data while varying configurational N, consistency score, and the structure of data has not been undertaken.


A principled test of the assessment \citet{lieberson_2004} described is useful in two ways. First, to generally assess the ability of QCA, with reasonable applications of existing robustness parameters (i.e. configurational N and consistency score thresholds), to filter random patterns. Second, it can be used for directly calculating the probability of returning random results according to unique features of any data set. This would be useful as an application-specific diagnostic assessment for QCA's ``truth table'' procedure, and would help protect researchers against wrongfully spotting patterns via stochastic properties of measurable phenomena. The interpretation of such an assessment would be the probability of returning a random configuration using the data and robustness thresholds selected by the user. Systematically applying Lieberson's suggested technique could also provide specific recommendations to inform researchers of an ideal consistency score and configurational N threshold to reach a desired level of confidence against a random result. 


Here, we determine whether QCA is robust to randomness, generally. We systematically apply QCA to thousands of random data sets, incrementally changing elements of the data structure -- sample size and the distribution of variables in the data set -- as well as elements under the control of the researcher -- consistency threshold, configurational N threshold and `complex' versus `parsimonious' solutions. We then use logistic regression to determine which of these elements affects the probability of returning a `random configuration,' or a result returned from random data. 

We also describe a related method, which is the primary purpose of this article. The Bootstrapped Robustness Assessment is to be used to evaluate any QCA result. This operates by generating many random data sets of the same data structure used in an application of QCA (i.e. based upon the sample size and variable distributions) and applying QCA repeatedly at the parameters under control of the researcher (i.e consistency score thresholds, configurational N, and parsimonious vs. complex solutions). The result is the probability that a given QCA application would return a random result, based upon random data of similar size and distribution. We hope that this method will provide straightforward, easily-interpreted recommendations for researchers who desire unarbitrarily-drawn parameters of choice.

\section{`Probabilistic Processes' and QCA}


A recent {\it{Sociological Methodology}} symposium on the methodological merits of QCA hosted a number of papers evaluating the reliability of QCA under both statistical and epistemological conditions. One paper saw problems with QCA under optimal conditions (\citealt{lucas_and_szatrowski_2014}) which subsequently could not be independently replicated by two other papers (\citealt{ragin_2014,vaisey_2014}). Seawright (2014) discusses the reliability of QCA when certain configurations are unobserved. One recent piece (\citealt{collier_2014}) points out several studies that verify whether QCA's algorithm returns robust configurations (e.g \citealt{hug_2013,schneider_and_wagemann_2012,krogslund_et_al_2015}). This paper will attempt to remedy what can be described as QCA's ``randomness problem'' -- the extent to which QCA's `truth table' analysis and algorithm are able to filter randomness in data given the robustness checks currently available. 

QCA's ``randomness problem'' is best described by \citet{lieberson_2004}. Lieberson's chief criticism is that ``QCA is less prepared to allow for chance and probabilistic processes'' than other methods and that ``procedures do not rule out the possibility that the observations are all a random matter'' (\citealt[13]{lieberson_2004}). Although QCA is a broad set of techniques to analyze small-to-medium-n data, the primary criticism of QCA has been its analysis of the `truth table.' The QCA `truth table' is a decomposition of data that analyzes each combination of causal conditions found in the data, the number of cases within each combination, and the extent to which cases that share these causal conditions agree on an outcome. An algorithm is applied to the truth table, combining the information into one or more ``causal recipes'' or solutions that result in an outcome. Lieberson imagines a test of his assertion that truth table analysis returns random results: apply QCA to a collected data set versus data where values arre randomly reassigned, keeping the marginal distributions intact. If QCA returns a configuration in both cases, it has a serious problem with being able to distinguish real patterns from random ones.

In a rebuttal, \citet{rihoux_and_ragin_2009} argue that such a test would show that random patterns would be filtered out by probabilistic procedures used in any application of QCA. One such procedure is the use of a high {\it{consistency threshold}}: the proportion of cases that are explained by a given a configuration or solution. This threshold is designed to prevent configurations that have high probability of being random from being included in the QCA algorithm and is the proportion of cases with the same combination of causal condition values and have the same value for the outcome. A recommended threshold is .85, meaning that 85\% of all cases with a specific configuration of causal conditions all agree in the value of the outcome. 

However, depending upon the marginal distributions of the conditions used in the QCA, a consistency score threshold could have a differential effect for filtering random configurations. For example, imagine an application of QCA had an outcome whose cases have a 90\% probability of being 1. Any combination of categories now has a .9 probability of having an outcome of value 1, with some rate of error. However, if attempting to explain the negation of the outcome, each combination of variables has a .1 probability of having an outcome with value 0. In these situations, the consistency score needed to protect the researcher against observing a random pattern differs quite a bit -- at base, a .9 is needed in the first case, while the lowest value is needed for predicting the negation. A more direct estimation of randomness, one that takes into account the marginal distribution of the outcome used in the analysis, would thus be helpful for providing an application-specific recommendation for the consistency score. 

A second probabilistic procedure to prevent spurious results is the {\it{configurational N threshold}}:  the minimum number of cases that have a certain combination of causal conditions, which allow the solution to be considered in the final result. This prevents the researcher from making conclusions about a small number of cases, especially about being overconfident about just one case with a unique set of conditions. To prevent such a scenario, the researcher can set a certain configurational N threshold (usually 2 or 3) to throw out those combinations of conditions that do not have a sufficient number of cases to make conclusions. 

Ostensibly, a high consistency score should be sufficient to account for error in causal conditions. As we have argued, these thresholds should have varying success rates according to the marginal probability of conditions present in the analysis. The utility of these procedures to distinguish a random data set from a collected one is an empirical question. Our first goal is thus to determine if the current researcher-set thresholds are enough to ensure that the results returned are not due to random chance. 

This paper addresses two areas of research inquiry. First, to what extent does a consistency score and configurational N threshold actually reduce the chance of returning a spurious result? Relatedly, how does this effectiveness differ according to the structure of the data? For example, is a consistency score of .9 effective at all sample sizes? How does the distribution of the outcome affect the usefulness of a high configurational N threshold? Our first set of results demonstrates the general effects of researcher choice on spuriousness in QCA's `truth table' analysis, with a special attention given to its variation with data structure. 

Our second set of questions refers to a practical application of these results. If results are highly dependent upon the structure of data analyzed, an application-specific robustness check would be useful for providing specific recommendations for researcher choice. The second set of results is a practical application of our procedure that uses a given QCA model, simulates many random data sets from this model, and 1) gives specific recommendations for ensuring against random results and 2) gives a specific value for the probability that a given QCA application would return a random result. The latter application has an analogous interpretation as a p-value for a QCA result. 





\section{Is QCA Robust to Randomness?}


There are researcher-set thresholds available prior to an analysis to reduce random configurations from being returned. The {\it{consistency threshold}} restricts the analysis to only consider configurational categories that have a certain proportion of cases that all agree in the direction of the dependent variable. The {\it{configurational N threshold}} restricts the analysis to only consider those configurational categories that have a certain number of cases within them. For example, we can choose to only include those combinations of causal conditions that have four or more cases that share all of the same causal condition states.

Though these are attempts to introduce probabilistic checks for QCA configurations, their use is often flexible, and general recommendations for which thresholds are hard to determine without a principled test of their usefulness. This section assesses the relative importance of each probabilistic check for filtering out random configurations from being returned by QCA. 

\subsection{Assessing the Robustness of QCA}

We employ a straightforward assessment of QCA using simulations. First, we first simulate a random data set. Next, we apply QCA to this random data set, and record whether QCA returned a result at all from random data. If we discover that a result is returned, we know that QCA is returning a spurious result. We systematically vary several variables to determine which elements of data structure (marginal distribution of variables, number of causal conditions included in the model, and sample size) and features of researcher choice (consistency threshold, configurational N threshold, and complex versus parsimonious solutions) affect the probability of a spurious result.  

In these simulations, each causal condition is a dummy variable with a marginal distribution randomly varying between .1 to .9 probability of being ``1.'' Though variables vary between iterations, all variables have the same marginal distribution within each iteration. The number of causal conditions vary from one to six. In accordance with QCA's focus for small- to medium-n samples, the sample size of the random data set varies from 10 to 60. Between iterations, we systematically vary the configurational N threshold from one to six. We vary the consistency threshold from .5 to 1.

The resulting data set is 2.5 million cases, each case an iteration of this procedure. We employ logistic regression on the results, with the dependent variable being a 0-1 outcome of returning a configuration from random data. The independent variables are the elements of data structure and researcher choice listed above. The primary question here is, which factors, when altered, independently decrease the probability of returning a random configuration in QCA?

Secondly, we assess whether researcher choice differentially affects the probability of a result given the structure of the data. For example, does a high consistency score threshold filter out spuriousness across all sample sizes? Does a high configurational N threshold filter spuriousness given all marginal probabilities of causal conditions? This is tested using interaction effects between variables measuring researcher choice and variables measuring data structure. If interaction effects are substantial, it suggests that additional assessments that take into account data structure need to applied to QCA to ensure robustness.

\subsection{Results}

Generally, the choice of a high consistency score and high configurational N in QCA do reduce the probability of returning a spurious result. Their effectiveness, however, is dependent upon the basic structure of the data: the distribution of the dependent variable, the number of variables used in the analysis, and the sample size affect how probable a result is spurious. First, we describe the main effects of each of these in turn. Then, we discuss how the structure of the data interacts strongly with checks. 

\subsubsection{The Effects of Data Structure and Researcher Choice on Spuriousness}

\begin{figure}[htb!]
\begin{center}
\label{diagram}
  \centerline{\includegraphics[scale=.5]{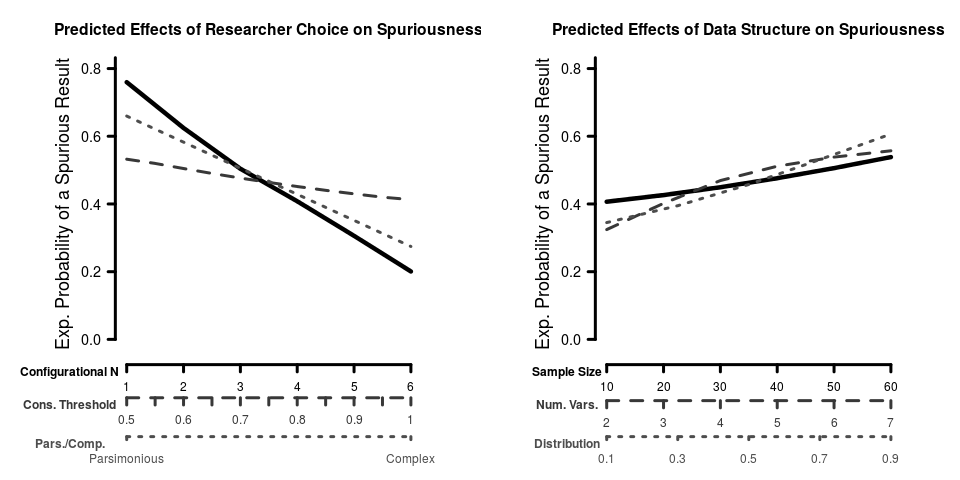}}
\vskip-1ex
\caption{Predicted Effects on the Probability of a Spurious Result in csQCA}
\end{center}
\end{figure}

Model 1 in Table 1 shows the results of the logistic regression model predicting whether QCA returned a spurious result from simulated random data. Figure 1 plots these effects graphically. The largest Model 1 effect is for the distribution of the outcome. An increase in the distribution of 1s in the dependent variable from probability 0 to probability 1 increased the logged odds of a spurious result by 2.29 (about 10 times the odds). This is an expected result -- in the case of an outcome with 90\% `1's', each case, and thus each configuration, has a 90\% of being 1 in the outcome. In Model 1, we see that increasing the number of variables decreases the probability of a spurious result by about 13\% for each additional variable. Interestingly, increasing the sample size also increases the odds of a spurious result by about 3\% for each additional case. This is in contrast to other mean-based approaches which typically become more robust as sample size increases.

\begin{table}[ht]
\caption{Logistic Regression of Elements of Researcher Choice and Data Structure Predicting a Spurious QCA Result }
\centering
\begin{tabular}{lll}
  \hline
Variable & Model 1 & Model 2 \\ 
  \hline
Intercept & 0.33(0.02)*** & 3.96(0.08)*** \\ 
  Cons. Threshold & -0.86(0.02)*** & -5.33(0.09)*** \\ 
  Conf. N Threshold & -0.71(0.00)*** & 0.08(0.01)*** \\ 
  Complex Solution & -0.82(0.01)*** & -1.41(0.12)*** \\ 
  Sample Size & 0.03(0.00)*** & -0.07(0.00)*** \\ 
  Num. Variables & -0.14(0.00)*** & 0.32(0.01)*** \\ 
  Dependent Variable Dist. & 2.29(0.01)*** & -5.88(0.07)*** \\ 
  Cons. Threshold * Conf. N Threshold &  & -0.57(0.02)*** \\ 
  Cons. Threshold * Complex Solution &  & -1.34(0.09)*** \\ 
  Cons. Threshold * Sample Size &  & -0.01(0.00)*** \\ 
  Cons. Threshold * Num. Variables &  & 0.23(0.01)*** \\ 
  Cons. Threshold * Dependent Variable Dist. &  & 10.11(0.08)*** \\ 
  Conf. N Threshold * Complex Solution &  & 0.01(0.01) \\ 
  Conf. N Threshold * Sample Size &  & 0.02(0.00)*** \\ 
  Conf. N Threshold * Num. Variables &  & -0.33(0.00)*** \\ 
  Conf. N Threshold * Dependent Variable Dist. &  & 0.27(0.01)*** \\ 
  Complex Solution * Sample Size &  & -0.04(0.00)*** \\ 
  Complex Solution * Num. Variables &  & 1.67(0.04)*** \\ 
  Complex Solution * Dependent Variable Dist. &  & -0.06(0.05)\\ 
  Sample Size * Num. Variables &  & 0.01(0.00)*** \\ 
  Sample Size * Dependent Variable Dist. &  & 0.05(0.00)*** \\ 
  Num. Variables * Dependent Variable Dist. &  & -0.37(0.01)*** \\ 
  AIC & 785374 & 670644 \\
   \hline
\multicolumn{3}{l}{\footnotesize{Null Deviance: 1029424 on 876083 degrees of freedom}}\\
\multicolumn{3}{l}{\footnotesize{$^{*}p < .05$, $ ^{**}p < .01$,  $ ^{***}p < .001$}}\\
\end{tabular}
\end{table}

Model 1 also shows the effect of researcher choice on returning a random configuration. Increasing the consistency score from and 0 to 1 in this case decreases the odds of returning a spurious result by 58\%. Using a ``complex'' versus ``parsimonious'' solution reduces the odds of a spurious result by 56\%. The most substantial effect, however, is the configurational N threshold, with a reduction in the odds of spuriousness by 51\% at each additional increase in the configurational N threshold. 

With the exception of two interactions (between complex solution and configurational N threshold, and between complex solution and the dependent variable distribution), Model 2 shows that researcher choice significantly moderates the effects of the structure of the data in the application of QCA. Because there are many strong interaction effects, we encourage the reader to focus upon the interpretation of Model 2. These variegated effects are described in the section and figures below.

\subsubsection{The Highly Contingent Effects of Researcher Choice on Spuriousness}

How effective is researcher choice in reducing the spuriousness of QCA under different conditions of data structure? One example question here is, To what extent does increasing the consistency score threshold effectively reduce the chance of a spurious result when the sample size is high?

\begin{figure}[htb!]
\begin{center}
\label{diagram}
  \centerline{\includegraphics[scale=.45]{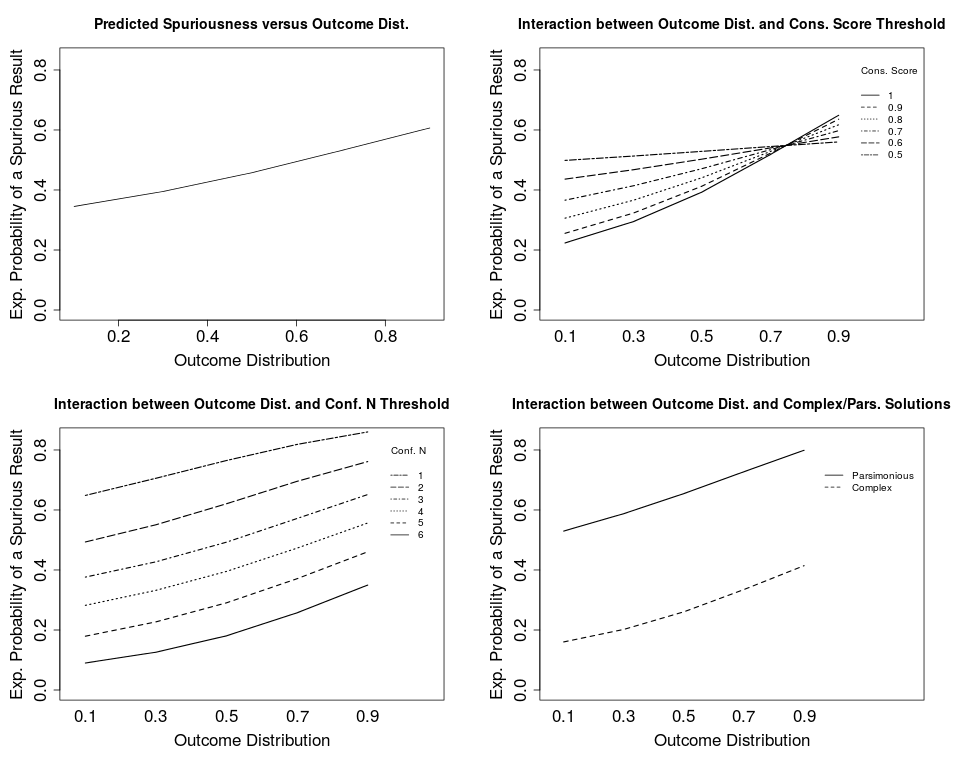}}
\vskip-1ex
\caption{Predicted Interactions between Elements of Researcher Choice and Outcome Distribution}
\end{center}
\end{figure}

Figure 2 shows the predicted interaction between researcher choice and the distribution of the dependent variable on the probability of spuriousness. On the y-axis is the expected probability predicted by the logistic regression model; the x-axis shows the outcome distribution in the data used to fit the QCA model. The top-left plot shows the predicted spuriousness across outcome distributions, with all other variables held at their means. The top-right shows that the consistency score threshold assists in reducing spuriousness by a great deal at lower distributions; when all cases have a .1 probability of being ``1'' on the outcome, increasing the consistency score from .5 to 1 decreases the probability of a spurious result from .5 to .2, a substantial reduction. The effectiveness of consistency score threshold on spuriousness decreases as the outcome distribution changes, however. When all cases have a .9 probability of being `1,' the predicted effect of increasing the consistency score is slightly negative. In data sets where cases almost all agree on an outcome, the consistency score may not be the most effective tool to prevent spuriousness. The higher likelihood that a lower consistency score improves robustness is likely due to lower consistency scores including all possible combinations into the analysis, which returns a non-result. Higher consistency scores in this circumstance will filter out configurations that randomly vary from the .9 baseline probability in the data set, returning a ``random'' configuration that made the cut while filtering out ``random'' configurations that predictably varied below the threshold in their incidence of the outcome.

The configurational N threshold substantially impacts the probability of robustness at all levels of the outcome distribution. Unlike the consistency score threshold, there is no change in the direction of the effect; increasing the configurational N threshold decreases the probability of spuriousness at all levels of the outcome distribution. The extent to the effect, however, decreases with an increase in proportion being `1' in the outcome. The probability of spuriousness at a configurational N threshold of 6 increases from .1 to .25 when the outcome distribution of `1s' increases from .1 to .9. 

As shown in the bottom-right plot in Figure 2, the ``complex'' solution is always more robust than the parsimonious solution, regardless of the outcome distribution. As shown in the Table 1, the effect is not significant. 

\begin{figure}[htb!]
\begin{center}
\label{diagram}
  \centerline{\includegraphics[scale=.45]{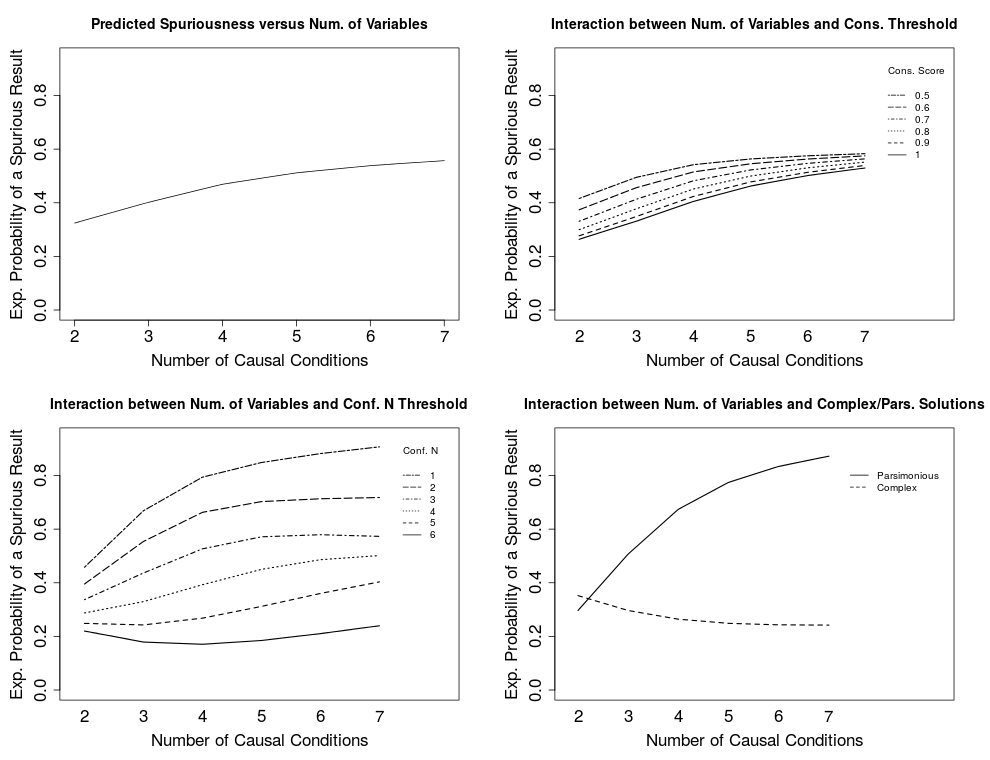}}
\vskip-1ex
\caption{Predicted Interactions between Elements of Researcher Choice and Number of Causal Conditions}
\end{center}
\end{figure}

Figure 3 shows the interactions between elements of researcher choice and the number of causal conditions used in the analysis. Generally, a larger number of causal conditions is predicted to have more spurious results. As the number of causal conditions increases, the effect of consistency score on spuriousness weakens. The opposite trend occurs when interacting configurational N threshold with the number of causal conditions -- the effect is much greater when using more conditions. Ostensibly, this is due to fewer configurations being included in the truth table analysis when the number of possible configurations increases. When the number of causal conditions is seven, there is a 60\% reduction in spuriousness when using a complex versus a parsimonious solution. At two causal conditions, however, the effect of using a complex versus parsimonious solution is negligible.

\begin{figure}[htb!]
\begin{center}
\label{diagram}
  \centerline{\includegraphics[scale=.5]{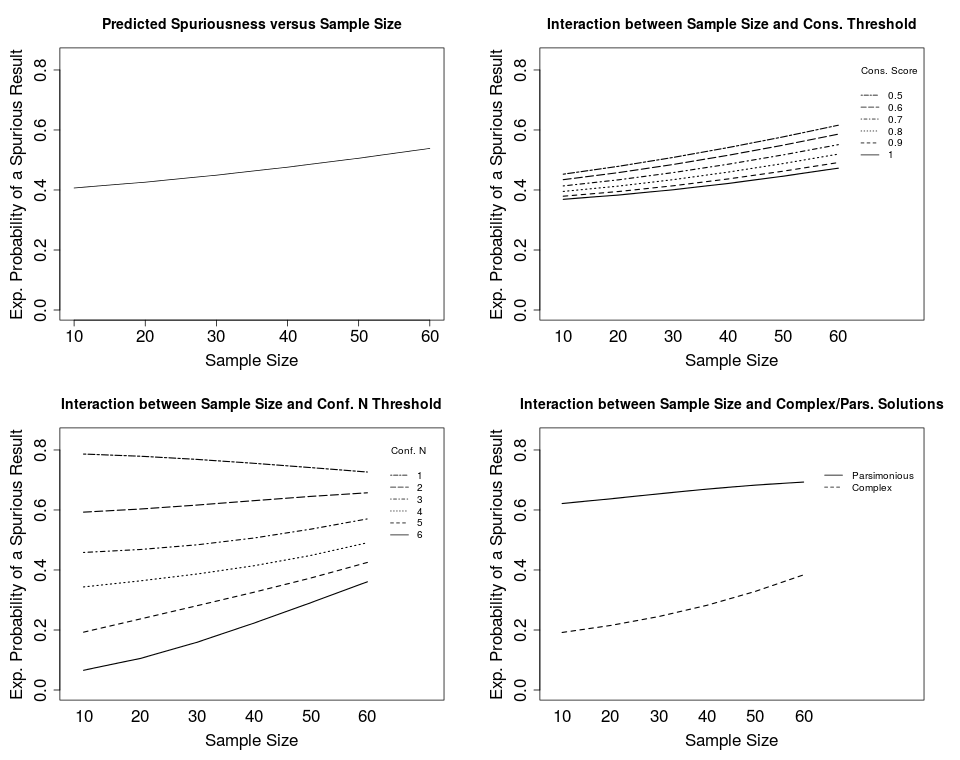}}
\vskip-1ex
\caption{Predicted Interactions between Elements of Researcher Choice and Sample Size}
\end{center}
\end{figure}

As Figure 4 shows, the effect of researcher choice varies the least when interacted with sample size. Table 1 shows significant effects for these interactions, the effect size is small compared to the interactions with outcome distribution and number of causal conditions.





\subsection{Discussion: QCA Robustness}

The results here show that the probabilistic checks set by the researcher in a QCA truth table analysis are effective in reducing the probability of a spurious result. However, they also show that the effectiveness of researcher-set parameters to ensure robust results vary according to the structure of the data. In some cases, consistency score alone will not effectively filter out random patterns, especially when the outcome is distributed such that all configurations have a high probability of having the same outcome value. In some cases, however, a high consistency score is not needed to ensure a robust result. The interaction effects above show that elements of researcher choice differed in their effectiveness at differing levels of number of causal conditions used. A high consistency score was more effective when using fewer causal conditions, while a high configurational N threshold was more effective at higher numbers of causal conditions. The difference between complex and parsimonious solutions was negligible at lower number of causal conditions, but represented a nearly four-fold increase in spuriousness -- from 25\% chance to 85\% chance -- when using seven causal conditions. 

These results show that the choices researchers make when conducting a truth table analysis of QCA data have a differential effect upon spuriousness according to variations of basic features of the data. The large variation of these effects, as well their complexity, justifies the need for a more straightforward approach to probabilistic assessments for QCA's truth table analysis. While it would be helpful to advise researchers on general practices for choosing a consistency score, a configurational N, and whether to use complex or parsimonious solution, the large variations in their effects according to features of the data used prevent the authors from doing so.

In the section below, we present a method for providing a model-by-model estimate of spuriousness. This method has two functions: 1) to estimate the 'confidence level' of an existing QCA model; and 2) to provide a reasonable recommendation for setting the consistency score and configurational N thresholds to achieve a desired `confidence level.' We first generally describe the method; we then give an example of its application 


\section{The Bootstrapped Robustness Assessment for QCA}

The Bootstrapped Robustness Assessment for QCA (baQCA) is a procedural check of a QCA result that takes into account data structure (e.g. marginal distribution of variables) and researcher choice (e.g. consistency score threshold) to provide an estimate of the probability of spuriousness for a given QCA result. 
Above, we show that the elements of researcher choice involved with ensuring robust QCA analysis of a truth table require vastly different thresholds according to the data structure. Identifying robust configurations would thus require taking into account the data structure.

To build a robustness assessment while taking into account data structure, we first draw a random data set using the same data structure as a QCA result. This includes using 1) the same number of causal conditions as the observed QCA data set, 2) the marginal distributions of the causal conditions and dependent variable present in the QCA data, and 3) the sample size. We then run a QCA model matching 4) the consistency score threshold set by the researcher and 5) the configurational N threshold set by the researcher. After thousands of repetitions, we take the simple probability that QCA returned a configuration given those parameters. The inverse of this proportion can be interpreted as the confidence that the configuration returned in the QCA analysis is due to random chance. This interpretation is similar to the p-value used in regression analysis to determine the ``significance'' of a result. Software to run this assessment is available using the R software package baQCA. 

Although there is current work being done to test the robustness of QCA to randomness (\citealt{braumoeller_2015,braumoeller_2015data}), this method randomly samples the outcome, keeping the causal conditions fixed, and permutes configurations rather than individual variables. Though QCA is at base a ``configurational method,'' we think it makes more sense to permute values of variables independent of other variables, rather than treating combinations of values for causal conditions as inextricably linked at predefined rates. Specifically, we argue that Braumoeller's focus on consistency score threshold as the primary check to prevent spurious results is inadequate, given our finding that the configurational N threshold is the much more powerful check. Though his method implicitly takes into account configurational N thresholds by including counts of configurations in his permutation model, these counts can vary considerably when variables are treated as independently sampled. Importantly, baQCA allows us to vary configurational N thresholds via independent random sampling of variables, and allows us to calculate how changes in configurational N threshold affect spuriousness. 

The idea behind this strategy is to protect the researcher from spurious results by estimating the probability of spuriousness given any random data. By repeatedly sampling random data over thousands of iterations, we are in effect directly observing this probability. When a 95\% confidence interval of the mean is calculated on a ``random'' variable, we conclude with 95\% confidence that the interval covers the true value of the mean. Our strategy identifies the probability that an application of QCA, with the exact data structure and parameters of researcher choice, would, with some level of confidence, return a result given completely random data. If this confidence level is low, the researcher should be cautious; when this confidence level is high, the researcher can conclude with confidence that the result is unlikely due to random chance, and is robust to a direct comparison with random data.

\subsection{baQCA in Practice}

This section outlines the step-by-step procedure of two methods for determining the probability that a given QCA application returns a spurious result. First, the Bootstrapped Assessment for QCA (baQCA) can be applied with a few steps: 

\begin{enumerate}
  \item `Fit' a QCA model with $v$ causal conditions and $n$ number of cases. 
  \item Simulate 2000 random data sets, each with $v$ causal conditions and size $n$. Each causal condition and the outcome has the same distributions as the observed data, respectively. 
  \item Apply QCA to all the generated data sets, matching the elements of researcher choice specified in the observed model (configurational N, consistency score, etc.). Record whether each QCA returned a spurious result. We chose 2000 to be the number of simulations using convergence diagnostics from \cite{gelman_and_rubin_1992}. 
  \item Take the simple proportion of times the QCA returned a configuration: 
  \begin{center}
  \[
  \frac{R}{2000} 
  \]
  {\it{Where $R$ = the number of times a QCA model returned a result from a randomly generated data set.}}
  \end{center}
  \item We use bootstrapped standard errors for a measure of uncertainty (\citealt{efron_and_tibshirani_1994}). We resample, with replacement, the vector of counts that sum to $R$. We chose 1000 resamples using convergence diagnostics from \citet{gelman_and_rubin_1992}. We then take 95\% quantiles of  $\frac{R}{2000}$ for our confidence interval.
\end{enumerate}

The resulting scalar, and its interval, is the 95\% confidence interval of the probability that the QCA application would return a random result. 

A useful, related tool is the {\it{recommendation}} of a consistency score/configurational N threshold given the data, without having fit a QCA model at all. To achieve this recommendation, we follow a similar trajectory: 

\begin{enumerate}
  \item Simulate 2000 random data sets, each with $v$ causal conditions and size $n$. Each causal condition and the outcome has the same distributions as the observed data, respectively. 
  \item Apply QCA to all the generated data sets, matching systematically varying parameters of researcher choice (configurational N, consistency score, etc.). Record whether each QCA returned a spurious result. Again, we chose 2000 simulations based on \cite{gelman_and_rubin_1992} convergence diagnostics. 
  \item We apply a logistic regression model to the results, using the configurational N and consistency score thresholds as predictors.
  \item We use this model to provide fitted values, calculating the minimum consistency score needed at every configurational N threshold to achieve a desired ``level of confidence.'' 
  \item We use the standard errors provided in the regression model to provide a confidence interval around each recommendation.
\end{enumerate}

Below, we provide a case study where reasonable thresholds were not quite enough to ensure good results, and where baQCA could be of service to suggest specific thresholds to ensure robustness.

\subsection{Qualitative Comparative Analysis of Tea Party Rallies in Florida}


In an application of this technique, we use a subset of data constructed by \citet{mcveigh_et_al_2014a} as part of their project on the emergence of Tea Party organizations in U.S. counties. The data set includes several county-level measures, including demographic measures from the American Community Survey (ACS) 2005-2009 (\citealt{acs_2009}), measures of religious adherence from the Association of Religion Data Archives (ARDA) 2001, 2008 Presidential election measures from Congressional Quarterly's {\it{America Votes}}, and the number of Tea Party organizations between 2009 and 2010 from the Institute for Research \& Education on Human Rights ((\citealt{irehr_2011})). We extend the data set to include the a new outcome variable, the number of rallies in each county between 2009 and 2010, also from IREHR. 

We restrict the data set to counties in Florida for two reasons based on our own case knowledge (\citealt{ragin_2008}). First, we choose Florida counties because, with the exception of California, all other states had fewer numbers of organizations. Second, and substantively important for the choice of causal conditions, we choose Florida due to the perceived impact the Tea Party movement had in the 2010 midterm election (\citealt{miller_and_walling_2012}). Restricting the data to counties in Florida leaves us with 67 cases for analysis (see Figure 5). Our analysis addresses the multiple causal pathways that lead to the occurrence of one or more Tea Party rallies in a Florida county.

\begin{figure}[htb!]
  \begin{center}
	\includegraphics[scale=.75]{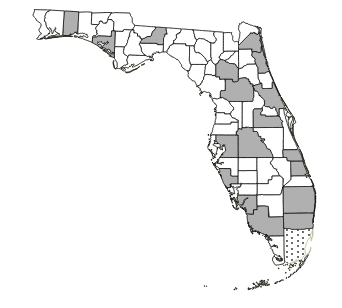} 
\vskip-1ex
\caption{Tea Party Rallies in Florida Counties. {\it{Note: Grey indicates at least one rally in county.}}}
\label{diagram}
\end{center}
\end{figure}

Qualitative Comparative Analysis arguments are combinational and often overlapping, and a researcher's causal conditions require a deep knowledge of cases in the data set for adequate placement within particular sets of causal conditions (\citealt{ragin_2008}). For example, to fully belong to a crisp outcome set, a Florida county must meet or exceed a minimum criterion. Therefore, it is necessary to both establish adequate causal conditions and an inclusionary criteria for each condition. Some have argued that because these criteria are based on researcher selection, criteria are biased (\citealt{lieberson_2004}), which can lead some to believe that a researcher has cherry-picked their analyses. To combat this assumption, and for the sake of clarity, we employ a simple inclusionary-exclusionary criterion for membership in a causal condition (outlined below) for the analysis. It is important to note that for many causal conditions, we dichotomize on the mean. Because QCA is designed to allow the researcher to be in dialogue between results and the cases and to recalibrate, it is generally considered bad practice to create inclusionary-exclusionary criteria in this way. We do so as a pedagogical exercise for the baQCA method and not as a manner with which to employ QCA.

Research on the Tea Party movement finds that while most of their organizations were concentrated in conservative partisan environments (\citealt{mcveigh_et_al_2014a,skocpol_and_williamson_2012}), much of their on-the-ground rally activity took place in heavily populated, left-leaning locales (\citealt{skocpol_and_williamson_2012,zernike_2010}). Research on Tea Party organizations has demonstrated the importance of educational background on support for the Tea Party (\citealt{mcveigh_et_al_2014a,skocpol_and_williamson_2012}), finding that supporters of the Tea Party movement are highly educated, and that Tea Party organizations were more likely to be established in U.S. counties characterized by a predominance of college graduates. Although supporters of the Tea Party movement tended to be relatively impervious to the economic recession of 2008 (\citealt{skocpol_and_williamson_2012,parker_and_barreto_2013}), many of the movement's grievances consisted of dismay about unemployment and the expanding reach of federal government, through a series of redistributive policies designed to remedy the economic situation (\citealt{skocpol_and_williamson_2012,mcveigh_et_al_2014a,parker_and_barreto_2013}). Scholars show that while much support for the Tea Party movement came from Protestants, these supporters were not of the Evangelical bent as was depicted by media (\citealt{zernike_2010,skocpol_and_williamson_2012,mcveigh_et_al_2014a}). Finally, \citet{parker_and_barreto_2013} argue that support for the movement derived from racial backlash against the nation's first black President, as well as fears that Barack Obama would initiate policies that would favor blacks. The size of the black population has two possible effects on support for the Tea Party. First, blacks pose a pose a potential threat in places where the black population is large because their predominance might encourage redistributive action by the Obama administration. Secondly, because rallies were much more likely to take place in left-leaning, densely-populated areas, these locales are also much more likely to have larger black populations.  

The brief summary of extant research on the Tea Party provides insight into creating causal arguments about the presence of Tea Party rallies as a test of QCA and the baQCA method. Importantly, because the analysis here employs crisp-set QCA, each causal condition is coded as either one or zero. As previously mentioned, the outcome variable is Tea Party RALLIES. Full placement in the outcome set (1) requires that a county has at least one rally. In sum, there are 19 cases that have the outcome. Given the differential relationship between Republican partisan contexts and the presence organizations and the occurrence of rallies, we include the measure REPUBLICAN. This is coded as one if, during the 2008 Presidential Election, the Republican candidate received a majority of the votes in the Florida county. Based on the above literature, we expect the absence of (negation of) Republican context as an important component of a causal pathway to Tea Party rallies. We also include four causal conditions in which full inclusion is defined in a straightforward manner: full membership in the causal set (1) is determined by whether a value for a particular case falls at or above the mean for that variable. 

First, we include a measure of COLLEGE educated, the percentage of people in the county (aged 25 or above) who hold a Bachelor's degree. Second, we include a measure of UNEMPLOYMENT, measured as the percent of the county population which is unemployed. With regard to membership in the college educated set, we expect that the presence of a college educated population is an important component of the pathway to rallies. However, given that many Tea Party supporters were not actually unemployed (although the movement's rhetoric says otherwise), we expect that the absence of a high unemployed population is an important part of explaining Tea Party rallies. Third, we include a measure for the size of the BLACK population, measured as the percentage of African-Americans in the county. Fourth, we include a measure for the size of the EVANGELICAL population, measure as the percentage of Evangelical adherents in the Florida county. 

In QCA, the logical representation of the presence of a causal condition is indicated by the variable name in all upper-case letters whereas negation is represented by all lower-case letters. Combinations of conditions (e.g. complex combinations of variables) in a pathway or recipe to an outcome are expressed as a string of variable names delineated by an asterisk, representing the logical operator ``AND.'' If multiple pathways exist, each pathway is delineated by a plus symbol, the logical operator for ``OR.'' Therefore, our main expectation, expressed in QCA notation, is:
\begin{center}
republican * COLLEGE * unemployment * BLACK * evangelical
\end{center}

In this analysis, there are a total of 32 possible pathways to the outcome, based on the five ($K$) causal conditions ($2^K = 2^5 = 32$). 

Before applying the baQCA method, we calculate Tea Party rally solutions based on plausible researcher preferences. In this test case, we use a default sufficiency inclusion or consistency score of .85 (the minimum proportion of cases explained by a causal configuration) and a default configurational N threshold of 1.0 (the minimum number of cases allowed for a particular configuration). The results in Table 2 indicate that only three configurations or combinations explain the presence of Tea Party rallies in Florida counties. Overall coverage for the three configurations is 47.4\%, with only BLACK appearing in each. The presence or absence of REPUBLICAN, COLLEGE, UNEMPLOYMENT, and EVANGELICAL appear in various combinations, known as ``insufficient but necessary components of causal combinations that are unnecessary but sufficient for the outcome'' (\citealt{mackie_1980,ragin_1987}). 

\begin{table}[h] 
\begin{center}
\caption{QCA Results for Florida Tea Party Rallies} \label{tab:title} 
\begin{tabular}{ >{$}l<{$}  >{$}c<{$} >{$}c<{$} >{$}c<{$}}
  \text{Solutions} &  & \text{Consistency} & \text{Coverage} \\
  \hline \hline
  & & & \\
  \text{COLLEGE * unemployment * BLACK * EVANGELICAL} &  & 100.0\% & 10.5\% \\
  \text{republican * COLLEGE * unemployment * BLACK} &  & 100.0\% & 36.8\% \\
  \text{republican * college * UNEMPLOYMENT * BLACK * evangelical} &  & 100.0\% & 5.3\% \\
  & & & \\
  \hline
\end{tabular}
\end{center}
\end{table}

\subsection{Application of the Bootstrapped Robustness Assessment}

As previously mentioned, using a high consistency score or altering the minimum number of cases for a solution is not enough to ensure the final configurations are non-spurious. To assess spuriousness in the current set of solutions, we apply the baQCA method, which yields a probability of randomness score; we also use the related irQCA method that suggests alternative consistency and configurational N thresholds, with significance values, for yielding a non-spurious set of solutions. 

Application of the baQCA method shows that causal configurations have a high probability of being random. In fact, of the simulated data sets, nearly 95 percent yielded a random result. The irQCA method suggests configurational N and consistency thresholds that would yield a non-spurious, ``significant'' result. The output is depicted in graphical form below. As shown and described above, to combat spuriousness in QCA, the researcher {\it{should}} select a high consistency threshold. In addition, the researcher should use their knowledge of the solutions and causal conditions to select an appropriate configurational N threshold. As shown in Figure 6, in all graphs except the bottom-right (p=.001) all suggested consistency thresholds reach or exceed .9. In addition, for these significance levels, the configurational N threshold ranges from three to five. Here, we use the smallest configurational N threshold, but the results are similar when using configurational N thresholds of three, four, and five. While it is possible to decrease the consistency score threshold and the configurational N threshold to yield additional configurations, these solutions would only cover one or two cases. Building configurations around one or two cases, instead of a medium-n set of cases, introduces randomness into the set of causal solutions and does not aid a researcher in developing substantive theory about the conditions that produce an outcome.

\begin{figure}[htb!]
\begin{center}
	\includegraphics[scale=.75]{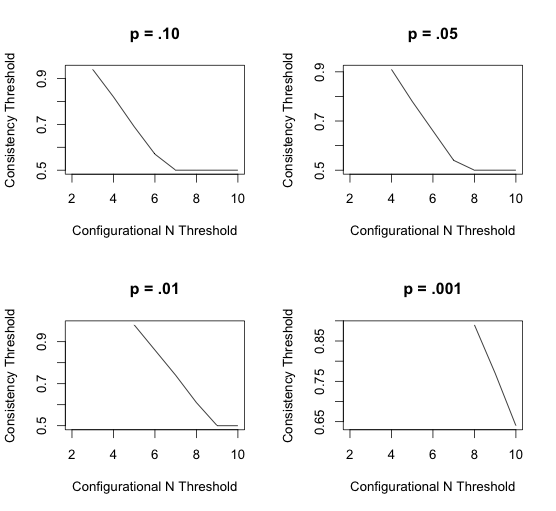} 
\vskip-1ex
\caption{Suggested Consistency and configurational N Thresholds by Desired Significance Level} 
\label{diagram}
\end{center}
\end{figure}

The selection of a minimum of three cases per solution, and 90 percent consistency results in one solution. This result is shown in Table 3 below. We can see that the result is exactly as expected by the literature above. Tea Party rallies were present in counties that were non-Republican, and had larger populations of college graduates and blacks, in combination with low levels of both unemployment and Evangelical adherents. This result is non-spurious at the .01 level. 

\begin{table}[h] 
\caption{QCA Results for Florida Tea Party Rallies, After Applying baQCA} \label{tab:title} 
\begin{center}
\begin{tabular}{ >{$}l<{$}  >{$}c<{$} >{$}c<{$} >{$}c<{$}}
  \text{Solutions} &  & \text{Consistency} & \text{Coverage} \\
  \hline \hline
  & & & \\
  \text{republican * COLLEGE * unemployment * BLACK * evangelical} &  & 100.0\% & 31.6\% \\
  & & & \\
  \hline
\end{tabular}
\end{center}
\end{table}

While the overall solution coverage decreased after applying baQCA, the subsequent set of solutions were less prone to randomness (checked by reapplying the baQCA method). The more robust configuration is a subset of the less robust configuration, suggesting a simpler solution that excludes from analysis the rows of the truth table that have high probability of spuriousness. Though it has been suggested to remove rows of the truth table with low coverage in general, this test provides a systematic way of doing so. As exhibited in Table 4, applying the thresholds recommended by the irQCA method improves our solutions. We are thus highly confident that the result returned by the QCA analysis is not due to randomness.

\begin{table}[h] 
\caption{QCA Results for Florida Tea Party Rallies, Difference by baQCA Application} \label{tab:title} 
\begin{center}
\begin{tabular}{ >{$}l<{$}  >{$}c<{$} >{$}c<{$}}
  \text{Solutions} & \text{Probability of Randomness} & 95\% \text{Confidence Interval} \\
  \hline \hline
  & & \\
  \text{Solution Set, before baQCA} & 0.9445 & 0.9360 \hspace{30pt} 0.9530 \\
  \text{Solution Set, after baQCA} & 0.0635 & 0.0550 \hspace{30pt} 0.0730 \\
  & & \\
  \hline
\end{tabular}
\end{center}
\end{table}

\subsection{Discussion: baQCA Results} 

The case above shows that a high consistency score threshold alone was not sufficient to prevent spuriousness from taking place. By increasing the configurational N threshold, we are able to greatly reduce the probability of spuriousness of the QCA result. In this case, the original consistency score was too conservative; the same level of significant is achieved by using a .9 consistency score threshold rather than 1. By estimating the probability of randomness via a direct comparison with random data, we are able to 1) estimate what thresholds are reasonable to achieve certain significance levels and 2) determine the probability of randomness of a QCA result.



\section{Conclusion and General Recommendations}

Though intuitive similarities between the Bootstrapped Assessment and p-values are obvious, we do not recommend using the same standard for QCA's truth table analysis as what is found in most p-value assessments (i.e. 95\% to 99\% confidence). The Bootstrapped Assessment, though a quantitative assessment of spuriousness, should be used to better inform the researcher about the conclusions drawn from the data rather than used as a black-and-white threshold of significance. This is to notify the researcher of whether the final result has an extremely high probability of including spurious configurations. If the method concludes that a QCA result is not robust to randomness, the researcher may need to more clearly articulate how the QCA results hold up, and are not due to random chance, based on their own case-oriented knowledge and analysis. 

Though QCA has been heavily criticized for not being robust to randomness, this simple assessment could help provide a systematic test of QCA against randomness. As we have shown, QCA can be robust to randomness using existing checks, but the ability of these checks to sufficiently protect against spuriousness differs according to data structure. We present a method that measures the confidence that a given QCA result is due to random chance, which provides information above and beyond the current robustness checks. Ideally, the Bootstrapped Robustness Assessment could be used as a standard assessment for QCA that provides precise recommendations for researcher choice.



\pagebreak{}


\bibliographystyle{asa_new}
\bibliography{library,ext_library}
\newpage

\end{document}